\documentclass[fleqn,usenatbib]{mnras}

\usepackage[T1]{fontenc}
\usepackage{ae,aecompl}

\usepackage{graphicx}	% Including figure files
\usepackage{amsmath}	% Advanced maths commands
\usepackage{atbegshi}

\usepackage{booktabs} 
\usepackage{rotating}

\usepackage{txfonts}
\usepackage{bm}
\title[Mechanical Model for Magnetized Blastwaves]{A Mechanical Model for Magnetized Relativistic Blastwaves}

\author[S. Ai and B. Zhang]{
Shunke Ai$^{1}$\thanks{E-mail: ais1@unlv.nevada.edu}
and Bing Zhang$^{1}$\thanks{E-mail: zhang@physics.unlv.edu}
\\
$^{1}$Department of Physics and Astronomy, University of Nevada Las Vegas, Las Vegas, NV 89154, USA\\
}

\date{Accepted XXX. Received YYY; in original form ZZZ}

\pubyear{2021}

\begin{document}
\label{firstpage}
\pagerange{\pageref{firstpage}--\pageref{lastpage}}
\maketitle

\begin{abstract}
The evolution of a relativistic blastwave is usually delineated under the assumption of pressure balance between forward- and reverse-shocked regions. However, such a treatment usually violates the energy conservation law, and is inconsistent with existing MHD numerical simulation results. A mechanical model of non-magnetized blastwaves was proposed in previous work to solve the problem. In this paper, we generalize the mechanical model to the case of a blastwave driven by an ejecta with an arbitrary magnetization parameter $\sigma_{\rm ej}$. We test our modified mechanical model by considering a long-lasting magnetized ejecta and found that it is much better than the pressure-balance treatment in terms of energy conservation. For a constant central engine wind luminosity $L_{\rm ej} = 10^{47}{\rm erg~s^{-1}}$ and $\sigma_{\rm ej} < 10$, the deviation from energy conservation is negligibly small at small radii, but only reaches less than $25\%$ even at $10^{19}{\rm cm}$ from the central engine. For a finite life time of the central engine, the reverse shock crosses the magnetized ejecta earlier for the ejecta with a higher $\sigma_{\rm ej}$, which is consistent with previous analytical and numerical results. In general, the mechanical model is more precise than the traditional  analytical models with results closer to those of numerical simulations.
\end{abstract}


%%%%%%%%%%%%%%%%%%%%%%%%%%%%%%%%%%%%%%%%%%%%%%%%%%

%%%%%%%%%%%%%%%%% BODY OF PAPER %%%%%%%%%%%%%%%%%%

\section{introduction}
When a relativistic ejecta powered by a central engine interacts with an ambient medium, a forward shock (FS) would propagate into the medium and a reverse shock (RS) would propagate into the ejecta. The fluid between the FS and RS is defined as a blastwave. Usually, an FS/RS system is divided into four regions: (1) unshocked ambient medium; (2) shocked ambient medium; (3) shocked ejecta; (4) unshocked ejecta. A contact discontinuity separates region (2) from region (3) \citep{sari95,zhang05}. 

Such a blastwave system is very relevant to the early phase of gamma-ray burst (GRB) afterglow emission. Particles accelerated from both FS and RS contribute to the observed afterglow emission \citep{meszarosrees99,saripiran99,zhang03,kobayashizhang03,wu03,mimica10}, see \cite{gao13} for a comprehensive discussion on all the possible spectral regimes and lightcurves from combined FS and RS emission. GRBs usually have a short duration so that the ejected shell has a finite thickness and RS shock crossing occurs around the blastwave deceleration radius \citep{sari95,zhang05}. In the existence of a long-lived central engine, e.g. a rapidly spinning pulsar or magnetar \citep{dailu98,zhangmeszaros01,ai18}, continuous injection of Poynting-flux energy would be possible. 

In the literature, an analytical treatment of an FS-RS blastwave system is usually  assumes pressure balance, i.e. $p_f = p_r$ where $p_f$ and $p_r$ are the pressure in the forward-shocked-region (region (2)) and reverse-shocked-region (region (3)), respectively. The Lorentz factor across the blastwave is roughly a constant in space, which is verified through hydrodynamical simulation \citep{kobayashi00}. This gives a reasonable, approximate treatment of the problem \citep{sari95,zhang05}. However, energy conservation is violated in such a treatment \citep{beloborodov06,yan07,uhm11}.
The reason is that pressure balance cannot be achieved immediately in a dynamically evolving system, and that there should exist a pressure gradient between the FS and RS. This is verified by the 1D MHD simulations conducted by \cite{mimica09} and their semi-analytical treatments, which derived scaling laws not attached to any particular effective thickness $\xi$ defined in \cite{sari95}, suggesting that pressure balance is generally not expected in relativistic blastwaves.
From the analytical perspective, \cite{beloborodov06} proposed a mechanical model to treat the problem more precisely, which breaks the pressure balance in the blastwave. The model was studied by \cite{uhm11} in detail, who demonstrated that energy conservation is preserved. In most of these treatments, a pure hydrodynamical (non-magnetized) blastwave was considered. 

Observations and theoretical modeling of GRB early afterglow \citep[e.g.][]{zhang03,troja17} and prompt emission \citep[e.g.][]{zhangyan11,yonetoku11,uhmzhang14} suggest that at least for some GRBs, the ejecta is magnetically dominated (see \citep{kumarzhang15} for a review). It is therefore interesting to study the RS dynamics for an arbitrarily magnetized relativistic outflow. A detailed analytical treatment of this problem was presented in \cite{zhang05} under the assumption of pressure balance (see also \cite{fan04} for the case of $\sigma<1$, \cite{giannios08} for a different analytical treatment and \cite{mimica09} for detailed numerical simulations). Denoting the magnetization parameter of the ejecta as $\sigma_{\rm ej} = B^2 / (4\pi \rho c^2)$, where $B$ is the magnetic field strength and $\rho$ is the mass density, both in the co-moving frame of the fluid. The pressure balance condition states $p_r + p_{r,b} = p_f$, where $p_f$ and $p_r$ are the gas pressures in the forward- and reverse-shocked regions, respectively, and $p_{r,b}$ is the magnetic pressure in the reverse-shocked region. Making use of the relativistic MHD shock jump condition \citep{kennel84,zhang05}, one can treat the evolution of the blastwave in detail. A criteria $\sigma_{\rm ej} < 8/3 \gamma_4^2 (n_1/n_4)$ for the formation of an RS was proposed based on the pressure balance assumption \citep{zhang05}\footnote{This condition was supported by the 1D Riemann problem solution by \cite{mizuno09}. \cite{giannios08} proposed that the RS shock should rather be $\sigma_{ej}\lesssim 0.02 \gamma_4^4\Delta^{3/2}(n_1/{\cal E})^{1/2}$, where $\Delta$ and ${\cal E}$ are the thickness and the energy of the ejecta, respectively). Detailed numerical simulations by \cite{mimica09} showed that a weak RS can exist in the regime where $\sigma$ is greater than this critical condition, but the rate of converting the total energy of the shell to heat is very low.}, where $n_1$ and $n_4$ are the number densities in regions (1) and (4), respectively, and $\gamma_4$ is the bulk Lorentz factor of the ejecta \citep{zhang05}. Such a treatment can roughly delineate the magnetized blastwave, especially when the central engine duration is short. However, the energy conservation condition is not satisfied, and the deviation becomes significant if the central engine powers a long-lasting magnetized wind. To treat such a problem, a mechanical model is desirable, but such a model does not exist in the literature for an arbitrarily magnetized outflow. 

In our work, we generalize the blastwave mechanical model to the regime for an ejecta with an arbitrary $\sigma_{\rm ej}$. In section \ref{sec:shock}, we review the basic criteria to excite a magnetized relativistic shock, the shock jump conditions and their solutions. In section \ref{sec:mechanical}, we derive the governing equations for the evolution of a magnetized blastwave in a mechanical model. In section \ref{sec:NS}, we present the results of a long-lived neutron star as the central engine as an example and test the energy conservation criterion. Conclusions are presented in section \ref{sec:con} with some discussion. 

\section{Magnetized relativistic shocks} 
\label{sec:shock}
In order to excite a shock in a relativistic hydrodynamic fluid, the relative speed between the two fluids should exceed the sound speed in the upstream, which reads \citep[e.g.][]{zhang18}
\begin{eqnarray}
c_s = c \sqrt{\hat{\gamma}p \over \rho_0 c^2 + {\hat{\gamma} \over \hat{\gamma}-1}p}
\end{eqnarray}
where $c$ is the speed of light, $\hat{\gamma}$ is the adiabatic index, which may be expressed as a function of the average internal Lorentz factor of the fluid \citep{kumar03,uhm11},
\begin{eqnarray}
\hat{\gamma} = {4{\bar{\gamma} + 1} \over 3\bar{\gamma}} .
\end{eqnarray}

For a magnetized fluid, one can define the magnetization parameter
\begin{eqnarray}
\sigma = {B_0^2 \over 4\pi \rho_0 c^2},
\label{eq:sigma}
\end{eqnarray}
where both $B_0$ and $\rho_0$ are the quantities in the comoving frame of the fluid. To excite a MHD shock in a magnetized ejecta, the relative speeds of two fluids must exceed the maximum speed of the fast magneto-acoustic (MA) wave in the upstream, which reads \citep[e.g.][]{leismann05,zhang18}
\begin{eqnarray}
v_{\rm F,max} &=& \sqrt{v_A^2 + c_s^2(1 - {v_A^2 \over c^2})} \nonumber \\
&=& c\sqrt{\hat{\gamma}p + {B_0^2 \over 4\pi} \over \rho_0 c^2 + {\hat{\gamma} \over \hat{\gamma}-1}p + {B_0^2 \over 4\pi}}
\end{eqnarray}
For a highly magnetized cold upstream, i.e. $\sigma \gg 1$ and $p \ll \rho_0 c^2$, the maximum speed of fast MA wave could be simplified and its corresponding Lorentz factor is
\begin{eqnarray}
\gamma_{\rm F,max} = \sqrt{1+\sigma}.
\end{eqnarray}

Once a shock is excited, the physical quantities in the upstream and downstream near the shock front are connected through the shock jump conditions. If the magnetic field lines are in the shock plane, the shock jump condition for a magnetized fluid reads \citep{kennel84,zhang05}
\begin{eqnarray}
n_1 u_{1s} & = & n_2 u_{2s} \\
E_s = \beta_{1s} B_{1s} & = & \beta_{2s} B_{2s}
\label{eq:Econs} \\
\gamma_{1s}\mu_1+{E_s B_{1s} \over 4\pi n_1 u_{1s}} & = &  \gamma_{2s}\mu_2 +  {E_s B_{2s} \over 4\pi n_2 u_{2s}} \\
\mu_1 u_{1s}+{p_1 \over n_1 u_{1s}} + {B_{1s}^2 \over 8\pi n_1 u_{1s}} & = & \mu_2 u_{2s}+{p_2 \over n_2 u_{2s}} + {B_{2s}^2 \over 8\pi n_2 u_{2s}},
\end{eqnarray}
where $n$ represents the particles' number density, $u = \gamma \beta$ is the four velocity in the direction of fluid's motion, 
\begin{eqnarray}
\mu = {h \over n} = m_p c^2 + e + p = m_p c^2 + {\hat{\gamma} \over \hat{\gamma}-1}{p \over n}
\label{eq:enthalpy}
\end{eqnarray}
is the specific enthalpy, $e$ is the internal energy density, and $p = (\hat{\gamma}-1)e$ is the thermal pressure. Here we adopt the convention that a quantity $Q_{ij}$ is defined as the value in region $i$ in the rest frame of $j$ and that the subscripts ``1'' and ``2'' represent the upstream and downstream, respectively, and the subscript ``s'' represents the shock. A quantity with only one subscript is defined in the rest frame of itself.  With the ``cold upstream" assumption, we have $p_1 = e_1 = 0$ and $\mu_1 = m_p c^2$. Notice that one has one additional jump condition for MHD shocks (Equation \ref{eq:Econs}) compared to the pure hydrodynamic shocks due to continuity of parallel electric field\footnote{Even though there is no electric field in the comoving frames of either upstream or downstream, in the rest frame of the shock (which moves relatively with respect to both streams) an electric field parallel to the shock front surface is induced due to Lorentz transformation, which is continuous across the shock. \ref{eq:Econs} is derived under the assumption that the plasma can be treated as a perfect conductor.}.

Noting $B_{is} = B_{i} {\gamma_{is}}$ ($i = 1,2$), using Equation \ref{eq:sigma} one can express the magnetization parameter in the upstream as 
\begin{eqnarray}
\sigma_1 = {B_1^2 \over 4\pi \rho_1 c^2} = {B_{\rm 1s}^2 \over 4\pi n_1 \mu_1 \gamma_{1s}^2}.
\label{eq:sigma_1}
\end{eqnarray}
Combining the jump conditions with Equations \ref{eq:enthalpy} and \ref{eq:sigma_1}, for a known $n_1$, all the quantities in the downstream can be expressed as a functions of $u_{2s}$, $\sigma_1$ and $\gamma_{21}$ \citep{zhang05}\footnote{Magnetic pressure $p_{b,i} = B_{i}^2 / 8\pi$ rather the strength of magnetic field was used in previous analyses. Here we consider $B_i$ directly for convenience of deriving the mechanical model later.}:
\begin{eqnarray}
u_{1s}&=&u_{2s}\gamma_{21}+(u_{2s}^2+1)^{1/2}(\gamma_{21}^2-1)^{1/2},\\
{n_2 \over n_1} & = & {u_{1s} \over u_{2s}} \\
{e_2 \over n_2 m_p c^2}&=&(\gamma_{21}-1)[1-{\gamma_{21}+1 \over 2 u_{1s}u_{2s}}\sigma_1] \\
{B_2 \over B_1} &=& {u_{1s} \over u_{2s}}.
\end{eqnarray}
Here $u_{2s}$ is calculated by solving a third-order equation derived from the jump conditions. Define $x = u_{2s}^2$, the equation reads \citep{zhang05}
\begin{eqnarray}
Jx^3 + Kx^2 + Lx + M = 0,
\label{eq:u2s}
\end{eqnarray}
where 
\begin{eqnarray}
J &=& \hat{\gamma}(2-\hat{\gamma})(\gamma_{21}-1) + 2, \\
K &=& -(\gamma_{21}+1)[(2-\hat{\gamma})(\hat{\gamma}\gamma_{21}^2+1)+\hat{\gamma}(\hat{\gamma}-1)\gamma_{21}]\sigma_1 \nonumber \\
&&-(\gamma_{21}-1)[\hat{\gamma}(2-\gamma)(\gamma_{21}^2 - 2)+(2\gamma_{21}+3)] \\
L&=& (\gamma_{21} + 1)[\hat{\gamma}(1-{\hat{\gamma} \over 4})(\gamma_{21}^2-1)+1]\sigma_1^2 \nonumber \\
&&+ (\gamma_{21}^2-1)[2\gamma_{21}-(2-\hat{\gamma})(\hat{\gamma}\gamma_{21}-1)]\sigma_1 \nonumber \\
&&+(\gamma_{21}-1)(\gamma_{21}-1)^2(\hat{\gamma}-1)^2 \\
M &=& -(\gamma_{21}-1)(\gamma_{21}+1)^2(2-\hat{\gamma})^2 {\sigma_1^2 \over 4},
\end{eqnarray}
with $\hat{\gamma} = (4\gamma_{21} + 1)/(3\gamma_{21})$.
Equation \ref{eq:u2s} can be solved numerically with a given $\sigma_1$ and $\gamma_{21}$. All the other quantities in the downstream right behind the shock front can be then calculated.

\section{A mechanical model for magnetized blastwaves}
\label{sec:mechanical}
\subsection{Ideal MHD equations}
Consider a magnetized FS-RS system which contains four regions. Instead of assuming pressure balance in the central two regions, we apply ideal MHD equations to describe the evolution of each fluid element. We have
\begin{eqnarray}
\nabla_{\mu}(\rho u^{\mu}) = 0
\label{eq:mass}
\end{eqnarray}
for mass conservation and
\begin{eqnarray}
\nabla_{\mu}T^{\mu \nu} = 0
\label{eq:T}
\end{eqnarray}
for energy-momentum conservation, where $\rho$ is the mass density of the blastwave in its comoving frame, $\mu_u$ is the normalized 4-velocity of the blastwave, and $T^{\mu \nu}$ is the energy-momentum tensor. For a magnetized blastwave, the energy-momentum tensor includes both fluid and electromagnetic components, i.e.
\begin{eqnarray}
T^{\mu \nu} = T^{\mu \nu}_{\rm FL} + T^{\mu \nu}_{\rm EM},
\end{eqnarray}
where 
\begin{eqnarray}
T^{\mu \nu}_{\rm FL} = (\rho c^2 + e + p) u^{\rm \mu} u^{\nu} + p \eta^{\mu \nu},
\end{eqnarray}
and 
\begin{eqnarray}
T^{\mu \nu} = {1 \over 4\pi} (F_{\lambda}^{\mu} F^{\lambda \nu} - {1\over 4} \eta^{\mu \nu} F^{\lambda \delta} F_{\lambda \delta}).
\end{eqnarray}
Here $e$ and $p$ stand for the internal energy and thermal pressure, and $F^{\mu \nu}$ is the electromagnetic tensor.

Explicitly splitting equation \ref{eq:mass} and \ref{eq:T} in 3+1 space-time, the dynamics of the blastwave can be delineated by the following ideal MHD equations \citep[e.g.][]{zhang18}:
\begin{eqnarray}
&{\partial (\gamma \rho) \over \partial t} + \nabla \cdot (\gamma \rho {\bf v}) = 0,  
\label{eq:mass_cons} \\
&{\partial \over \partial t}({\gamma^2 h \over c^2}{\bf v} + {{\bf E_L} \times {\bf B_L} \over 4 \pi c}) + \nabla \cdot [{\gamma^2 h \over c^2}{\bf v} \otimes {\bf v} + (p + {E_L^2 + B_L^2 \over 8\pi}){\bf I} \nonumber\\
&- {{\bf E_L} \otimes {\bf E_L} + {\bf B_L}\otimes {\bf B_L} \over 4\pi}] = 0,
\label{eq:momentum_cons} \\
&{\partial \over \partial t}(\gamma^2 h - p -\gamma \rho c^2 + {B_L^2 + E_L^2 \over 8\pi}) \nonumber \\
&+\nabla \cdot [(\gamma^2 h - \gamma\rho c^2){\bf v} + {c \over 4\pi}{\bf E_L} \times {\bf B_L}] = 0.
\label{eq:energy_cons}
\end{eqnarray}
Here $B_L$, $E_L$ and $v$ are the quantities defined in the lab frame, while others are in the rest frame of the fluid.  
Considering that the plasma in the blastwave can be treated as a perfect conductor, one can derive the strength of electric field as 
\begin{eqnarray}
{\bf E_L} = - {{\bf v} \over c} \times {\bf B_L} = - \bm{\beta} \times {\bf B_L}.
\end{eqnarray}

\subsection{Governing equations for the evolution of the blastwave}
\label{sec:evolution}

Since astrophysical blastwaves are usually powered by a point-source central engine, we consider a spherical geometry ($r$,$\theta$,$\phi$) throughout the paper. Since the ambient medium is usually not highly magnetized, we consider the interaction between a magnetized ejecta and a non-magnetized medium.

To simplify the ideal MHD equations, we assume that the magnetic field lines in region 4 are in the $\phi$ direction, which is parallel to the shock plane. Shock jump conditions dictate that the magnetic field lines in region 3 have the same direction as that in region 4. The bulk motion direction of the blastwave is in the radial direction, i.e. ${\bf v} = v {\bf e_r}$ so the electric field direction in the blastwave as viewed in the lab frame is in the $\theta$ direction, i.e. ${\bf E_L} = E_L {\bf e_\theta} = \beta B_L {\bf e_\theta}$. Therefore, we have 
\begin{eqnarray}
{\bf E_L} \times {\bf B_L} = \beta B_L^2 {\bf e_r},
\end{eqnarray}
\begin{eqnarray}
{\bf E_L} \otimes {\bf E_L} = \left[\begin{array}{ccc}
0 & 0 & 0 \\
0 &E_L^2 & 0 \\
0 & 0 & 0 
\end{array} \right] = 
\left[\begin{array}{ccc}
0 & 0 & 0 \\
0 &\beta^2 B_L^2 & 0 \\
0 & 0 & 0 
\end{array} \right] 
\end{eqnarray}
and
\begin{eqnarray}
{\bf B_L} \otimes {\bf B_L} = \left[\begin{array}{ccc}
0 & 0 & 0 \\
0 & 0 & 0 \\
0 & 0 & B_L^2 
\end{array} \right].
\end{eqnarray}
With $B_L = \gamma B$ (where $B$ is the magnetic field of blastwave in its rest frame),  Equations \ref{eq:momentum_cons} and \ref{eq:energy_cons} can be simplified as

\begin{eqnarray}
&&{1 \over c}{\partial \over \partial t}(\gamma^2 h \beta) + {1 \over 4\pi c}{\partial \over \partial t}(\gamma^2 \beta B^2) + {1 \over r^2}{\partial \over \partial r}(r^2 \gamma^2 h \beta^2) + {\partial p\over \partial r} \nonumber\\
&& + {(1+\beta^2) \over 8\pi}{\partial  \over \partial r}( \gamma^2 B^2) + {1 \over 4\pi r}(1+\beta^2)\gamma^2 B^2 = 0
\label{eq:momentum2}
\end{eqnarray}
and
\begin{eqnarray}
&&{\partial \over \partial t}(\gamma^2 h) - {\partial \over \partial t}p + {1 \over 8\pi}{\partial \over \partial t}[(1+\beta^2)\gamma^2 B^2] \nonumber\\
&&+{1 \over r^2} {\partial \over \partial r}(r^2 \gamma^2 h \beta c) + {c \over 4\pi r^2}{\partial \over \partial r}( r^2\beta \gamma^2 B^2) = 0.
\label{eq:energy2}
\end{eqnarray}

Instead of investigating the profiles of various quantities in the blastwave, we define some integrated variables:
\begin{eqnarray}
 \Sigma & = & \int_{r_r}^{r_f} \rho dr, \label{eq:Sigma}\\
 P & = & \int_{r_r}^{r_f} p dr, \\
 H & = & \int_{r_r}^{r_f} h dr, \\
 {\cal B} & = & \int_{r_r}^{r_f} B^2 dr, \label{eq:calB}
\end{eqnarray}
where $r_r$ and $r_f$ represent the distances of RS and FS from the central engine. Notice that the first three integrals were defined in the original mechanical model \citep{beloborodov06,uhm11}. Also, we keep the assumption a constant velocity in the blastwave so that ${\partial \beta \over \partial r} = 0$. 
Notice an identity for any function $f(t,r)$
\begin{eqnarray}
\int_{r_r(t)}^{r_f(t)} {\partial \over \partial t}f(t,r) dr &=& {d \over dt} \left[\int_{r_r}^{r_f} f(t,r) dr \right] \nonumber\\
&+& c[f_r \beta_r - f_f \beta_f],
\end{eqnarray}
where $f_r$ and $f_f$ are the values of $f$ right behind the RS and FS in the rest frame of the blastwave, respectively, and $\beta_r$ and $\beta_f$ are the velocities of RS and RS in the lab frame, respectively. One can then integrate Equations \ref{eq:mass_cons}, \ref{eq:momentum2}, and \ref{eq:energy2}. Define the distance of the contact discontinuity from the central engine as $r_d$ and the dimensionless speed of the contact continuity as $\beta$, one then has ${d\over dt} = \beta c {d \over d r_d}$. The three equations can be then expressed as\footnote{Notice that we do not consider the profiles of the quantities in the blastwave. Rather, we approximate the defined integrated quantities as the properties of a point-like fluid at contact discontinuity $r_d$, i.e. $f(r)=F\delta(r-r_d)$, where $F$ stands for any of the integrated quantities defined in Equations \ref{eq:Sigma} - \ref{eq:calB}. However, there should be no time derivative involved in the terms with this approximation.}

\begin{eqnarray}
{\beta \over r_d^2}{d \over dr_d}(r^2 \Sigma \Gamma) = \Gamma [\rho_r(\beta-\beta_r)+ \rho_f(\beta_f - \beta)]  
\label{eq:mass3}
\end{eqnarray}

\begin{eqnarray}
&&{\beta \over r_d^2} {d \over dr_d}(r^2\Gamma^2 H \beta)  - \Gamma^2 \beta [h_r(\beta-\beta_r)+ h_f(\beta_f - \beta)] \nonumber\\
&+& {\beta \over 4\pi }{d \over dr_d}(\Gamma^2 \beta {\cal B}) + {\beta \Gamma^2 \over 4\pi}[ B_r^2 \beta_r -  B_f^2 \beta_f]+ (p_f - p_r) \nonumber\\
&+& {\Gamma^2 (1+\beta^2) \over 8\pi}(B_f^2 - B_r^2) + {(1+\beta^2)\Gamma^2{\cal B} \over 4\pi r_d} = 0
\label{eq:momentum3}
\end{eqnarray}

\begin{eqnarray}
&&{\beta \over r^2} {d \over dr_d}(r^2\Gamma^2 H) -\Gamma^2 [h_r(\beta-\beta_r) + h_f(\beta_f - \beta)] \nonumber \\
&-& \beta {dP \over dr_d} - (\beta_r p_r - \beta_f p_f )+{\beta \over 8\pi}{d \over dr_d}[(1+\beta^2)\Gamma^2{\cal B}] \nonumber \\ 
&+&{(1+\beta^2)\Gamma^2 \over 8\pi}(\beta_r B_r^2 - \beta_f B_f^2)+{\Gamma^2 \beta \over 4\pi}(B_f^2 - B_r^2)  \nonumber\\
&+& {\beta \Gamma^2 {\cal B} \over 2\pi r_d} = 0
\label{eq:energy3}
\end{eqnarray}
Here, $\Gamma = \gamma$, which is used to keep consistency with the format of other variables. Since ${d\beta \over dr_d} = {1 \over \beta \gamma^3} {d\gamma \over dr_d}$, we totally have 5 independent unknowns ($\Gamma$, $\Sigma$, $P$, $H$, ${\cal B}$). Besides Equations \ref{eq:mass3} - \ref{eq:energy3}, one needs two more equations to close the problem. The first one is the equation of state of the fluid, which reads \citep[e.g.][]{beloborodov06,uhm11}.
\begin{eqnarray}
H = \Sigma c^2 + {\hat{\gamma} \over \hat{\gamma} - 1}P.
\label{eq:eos}
\end{eqnarray}
Another equation comes from the accumulation of ${\cal B}$ during the propagation of the reverse shock in the ejecta.
Practically, it is easier to calculate an integral over volume than over radius. 
Define ${\cal B}_{\rm sph} = \int B^2 dV$, where $dV = dV'/\Gamma$ is the incremental volume at the RS in the lab frame and $dV'$ is that in the comoving frame. The incremental particle number at the RS front is $dN = \rho_r dV' / m_p$, which is defined by the properties of the injected wind by
\begin{eqnarray}
dN = {dE_{\rm inj,p} \over \gamma_4 m_p c^2},
\label{eq:dN}
\end{eqnarray}
where $dE_{\rm inj,p}$ is the injected particle kinetic energy during the lab-frame time $dt$ into the RS.
Assuming that the magnetization parameter in each $dN$ shell is uniform, one can express the magnetization parameter at the RS downstream as
\begin{eqnarray}
\sigma_r = {B_r^2 dV \over 4\pi \rho_r c^2 (dV'/\Gamma)} = {d{\cal B}_{\rm sph} \over 4\pi m_p c^2 (dN/\Gamma)}. 
\label{eq:sigma_bw}
\end{eqnarray}
For a low-$\sigma$ relativistic blastwave, $r_r$ and $r_f$ are very close so that one may adopt the approximation $r_r \approx r_f \approx r_d$. However, in the high $\sigma$ regime, the RS velocity in the lab frame, $\beta_r$, is significantly smaller than the FS velocity in the lab frame, $\beta_f$. Under certain conditions, the RS could even move back towards the central engine. The $r_r \approx r_f \approx r_d$ approximation is no longer valid. Since $B$ may change much more drastically near the RS than anywhere else, the approximation ${dB^2 \over dt}(r) = \delta (r-r_r) \int_{r_r}^{r_f}{dB^2(r) \over dt}dr$ is taken. From the identity 
\begin{eqnarray}
&&{d \over dt}\left[\int_{r_r}^{r_f} f(r,t) dr \right] = \int_{r_r}^{r_f} {df(r,t) \over dt} dr \nonumber \\
&& + c[f_r(t)(\beta-\beta_r) + f_f(t)(\beta_f - \beta)]
\end{eqnarray}
for any $f(r,t)$, and the fact $B_f = 0$, the relation between ${\cal B}$ and ${\cal B}_{\rm sph}$ evolution may be written as
\begin{eqnarray}
{d{\cal B}_{\rm sph} \over dt} &=& {d \over dt}\left[\int_{r_r}^{r_f} 4\pi r^2 B^2 dr\right] \nonumber \\
&=&\int_{r_r}^{r_f}{d \over dt}(4\pi r^2B^2)dr + 4\pi r_r^2 B_r^2(\beta - \beta_r) c \nonumber \\
&\simeq& 8\pi r_d {\cal B}\beta c + 4\pi r_r^2 \int_{r_r}^{r_f} {dB^2 \over dt} dr \nonumber \\
&&+ 4\pi r_r^2 B_r^2(\beta - \beta_r) c \nonumber \\
&\simeq&  8\pi r_d {\cal B}\beta c + 4\pi r_r^2 {d {\cal B}\over dt}
\label{eq:Bsph}
\end{eqnarray}
Rewriting Equation \ref{eq:Bsph} in terms of $dr_d$ instead of $dt$, one gets
\begin{eqnarray}
{d{\cal B} \over dr_d} = {1 \over 4\pi r_r^2}{d{\cal B}_{\rm sph} \over dr_d} - 2{\cal B}{ r_d \over r_r^2},
\label{eq:dB}
\end{eqnarray}
where $d{\cal B}_{\rm sph}$ can be obtained from Equations \ref{eq:dN} and \ref{eq:sigma_bw}, once $dE_{\rm inj}$ is given. Now we have closed the problem. The evolution of the blastwave is governed by Equations \ref{eq:mass3} - \ref{eq:energy3}, \ref{eq:eos} and \ref{eq:dB}.

\section{Blastwave powered by a long-lasting magnetized ejecta}
\label{sec:NS}
We now apply the mechanical model to study the dynamics of a blastwave powered by a long-lasting magnetized ejecta with a constant magnetization parameter $\sigma_{\rm ej}$. It interacts with an ambient medium to excite an FS - RS system under some conditions\footnote{The criteria $\sigma_{\rm ej} < (8/3)\gamma_4 (n_1 / n_4)$ proposed in \cite{zhang05} is a good approximation in most cases. In this paper, we only use the most fundamental criterion, which requires the relative speed of two fluids to be greater than the sound speed (or maximum speed of the fast MA wave) in the upstream fluid.}. For simplicity, we assume a constant Lorentz factor ($\gamma_{\rm ej}$) for the ejecta. Then the energy injected into the blastwave in the lab frame at each lab time interval ($dt$) can be calculated as
\begin{eqnarray}
dE_{\rm inj} = L_{\rm inj} dt = L_{\rm inj} {dr_d \over \beta c}
\end{eqnarray}
where $L_{\rm inj}$ is the luminosity of energy injection in the lab frame. Consider a shell with energy $dE_{\rm inj}$ that was ejected from the central engine in a engine time interval $d\tau$. Then, the luminosity of central engine can be written as $L_{\rm ej} = dE_{\rm inj}/d\tau$. Considering two thin fluid layers ejected from the central engine at $\tau_1$ and $\tau_2$, which would reach the RS at $t_1$ and $t_2$, one has 
\begin{eqnarray}
\beta_{\rm ej}(t_1-\tau_1) + \beta_r (t_2 - t_1) = \beta_{\rm ej}(t_2 - \tau_2),
\end{eqnarray}
from which $\beta_{\rm ej} d\tau = (\beta_{\rm ej} - \beta_r) dt$ can be derived. Hence, 
\begin{eqnarray}
L_{\rm inj} = L_{\rm ej}{\beta_{\rm ej}-\beta_r \over \beta_{\rm ej}}
\end{eqnarray}
The contribution from the injected particle kinetic energy to the total injected energy is $dE_{\rm inj,p} = dE_{\rm inj}/(1+\sigma_{\rm ej})$, which is used to calculate $dN$ in Equation \ref{eq:dN}.

With known $L_{\rm ej}$ and $\sigma_{\rm ej}$, one can calculate the quantities in region 4 near the RS, including the number density
\begin{eqnarray}
n_4 = {L_{\rm ej} \over 4\pi r_r^2 \beta_{\rm ej} \gamma_{\rm ej}^2 c^3 m_p (1+\sigma_{\rm ej})}
\end{eqnarray}
and the magnetic field 
\begin{eqnarray}
B_4 = (4\pi n_4 m_p c^2 \sigma_{\rm ej})^{1/2}.
\end{eqnarray}
Given an initial value of $\Gamma$, one can calculate the relative velocity between the bulk motion of the blastwave and the unshocked ejecta, which reads
\begin{eqnarray}
\beta_{34} = {\beta_{\rm ej} - \beta \over 1 - \beta \beta_{\rm ej}}.
\end{eqnarray}
so that the corresponding Lorentz factor is $\gamma_{34} = [1/(1-\beta_{34}^2)]^{1/2}$. Now we can solve Equation \ref{eq:u2s} to solve $u_{3,rs}$ and then calculate $\rho_r$, $p_r$, $h_r$ and $B_r$. Similarly, we have $\gamma_{\rm 21} = \Gamma$ and obtain $\rho_f$, $p_f$, $h_f$ and $B_f$. Note that $B_f = 0$ is satisfied for non-magnetized ISM. Substituting the values of the quantities at the forward and reverse shocks to the governing equations listed in \ref{sec:evolution}, the evolution of the blastwave can be solved.

\begin{figure*}
\begin{center}

\begin{tabular}{ll}
\resizebox{80mm}{!}{\includegraphics[]{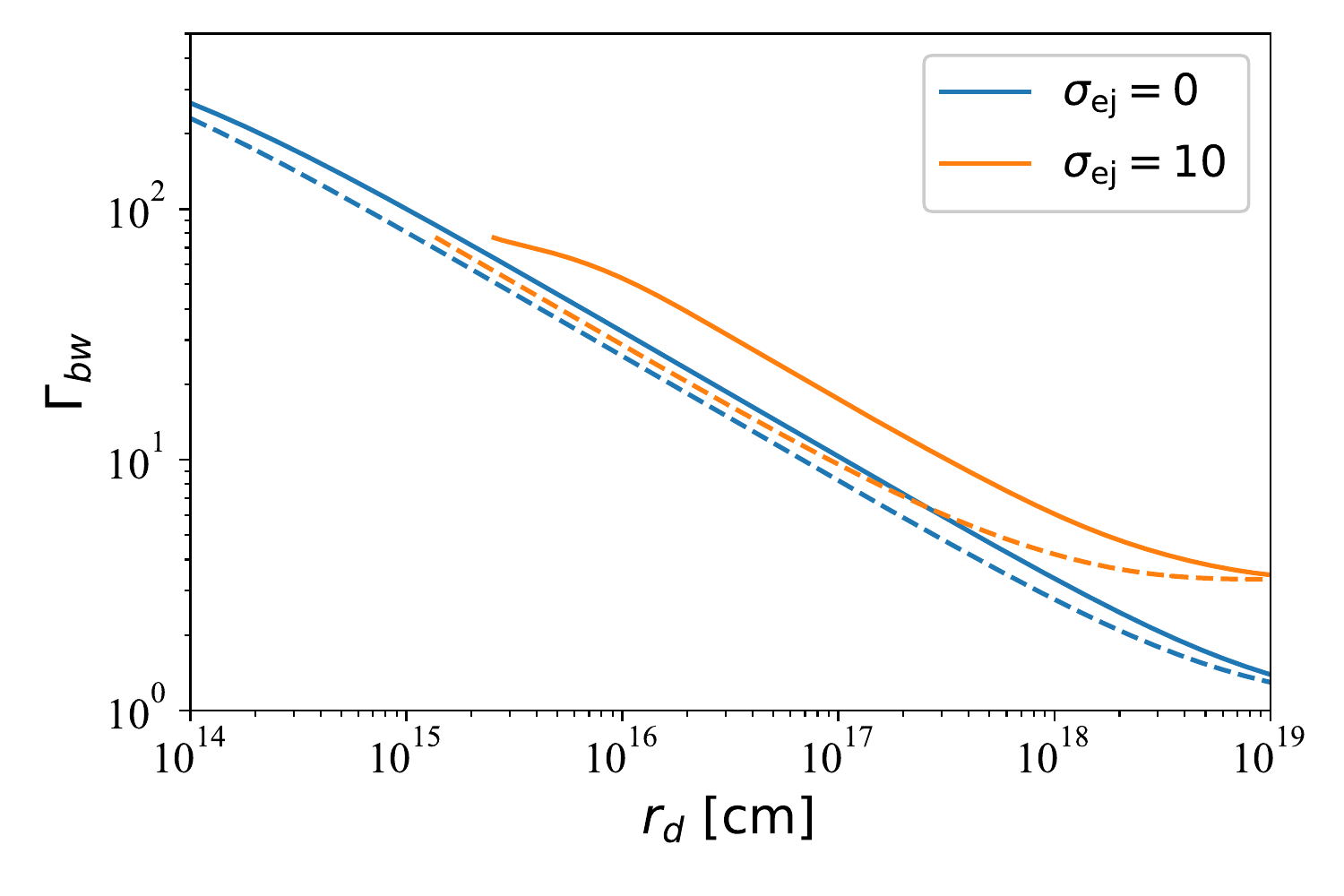}}&
\resizebox{80mm}{!}{\includegraphics[]{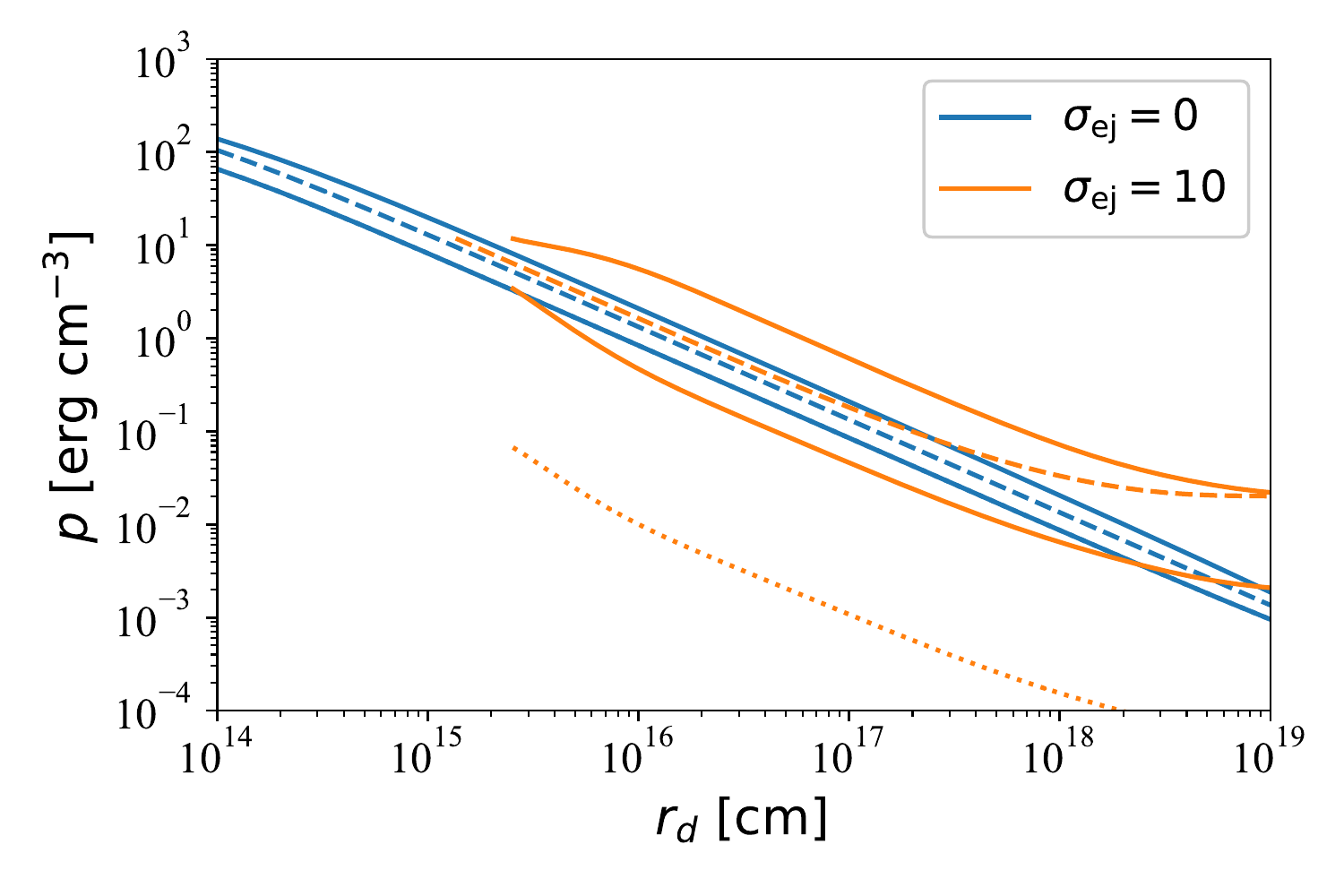}}
\\
\resizebox{80mm}{!}{\includegraphics[]{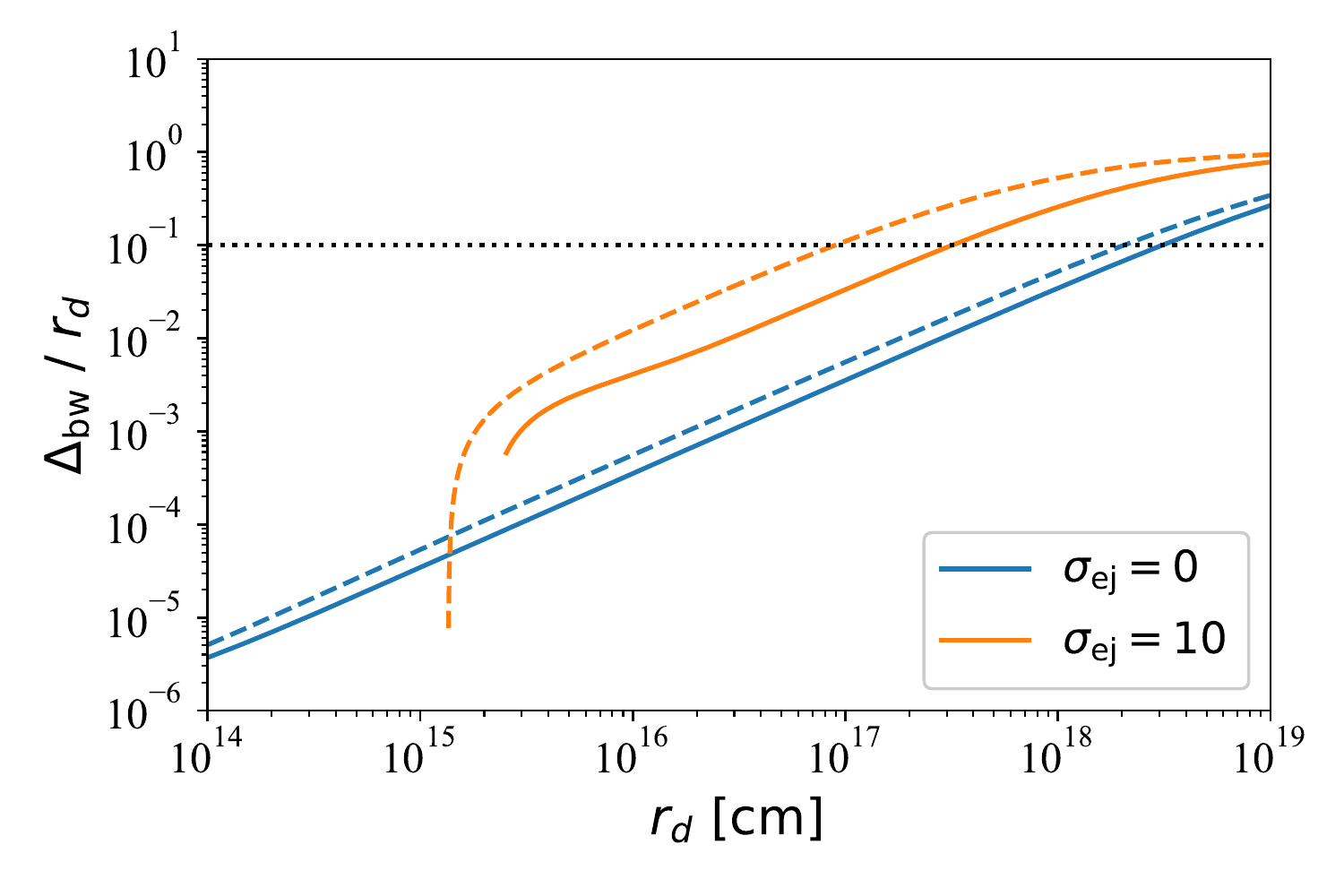}}&
\resizebox{80mm}{!}{\includegraphics[]{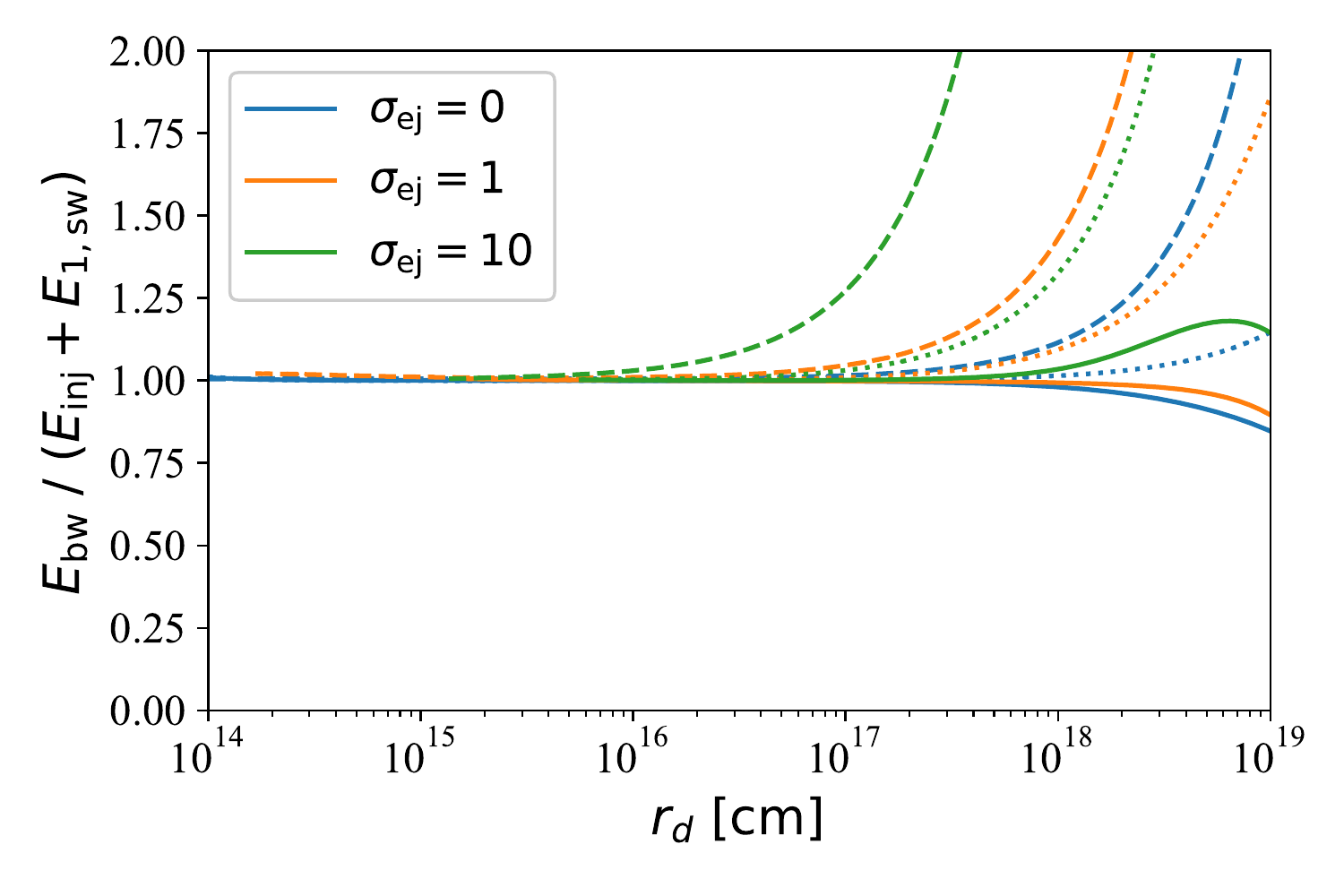}}
\end{tabular}
\caption{The evolution of the properties of the blastwave with $L_{\rm ej} = 10^{47}{\rm erg~s^{-1}}$ and an infinite central engine lifetime. $\gamma_{\rm ej} = 500$ and $n_1 = 1 {\rm cm^{-3}}$ are assumed. Different colors represent different values of the magnetization parameter $\sigma_{\rm ej}$. Solid lines represent the mechanical model and dashed lines represent the pressure balance model. Upper left panel: the evolution of Lorentz factor of the blastwave; Upper right panel: the evolution of pressure. The solid lines above and below the dashed lines represent total pressure behind RS ($p_{\rm r,tot}$) and FS ($p_f$) respectively. The dotted dashed line is the thermal pressure behind the reverse shock ($p_r$) with $\sigma_{\rm ej} = 10$. Lower left panel: the thickness of the blastwave normalized to the radius of contact discontinuity.The black dotted line represents the level where the thickness is an order of magnitude smaller than the radius of contact discontinuity. Lower right panel: The ratio between the blastwave's energy and the energy injected to the blastwave from the RS and FS. We calculate the energy of blastwave through Equation \ref{eq:Ebwp} for the pressure balance model (dashed lines) and Equation \ref{eq:Ebwm} for the mechanical model (solid lines). We also calculate the blastwave energy for the mechanical model with Equation \ref{eq:Ebw} (dotted lines).}
\label{fig:evolving}
\end{center}
\end{figure*}

Figure \ref{fig:evolving} shows the calculated blastwave evolution in the mechanical model. For comparison, we also plot the evolution of the blastwave under the pressure balance assumption in the same figure. As one can see, there is an apparent difference between the pressure balance model and the mechanical model. It has been discussed in \cite{uhm11} that once pressure balance was assumed, the expansion of the blastwave caused by $pdV$ work would be ignored, which would lead to an underestimation of the blastwave's Lorentz factor $\Gamma$. In the magnetized blastwaves, the contribution of magnetic pressure is equivalent to thermal pressure. Hence, $\Gamma$ is again  underestimated for a magnetized fluid in the pressure-balance model.

We test the mechanical model from the view point of energy conservation. The total energy of the blastwave can be obtained by integrating the 00 component of the energy-momentum tensor over the volume between the forward and the reverse shocks, which is expressed as
\begin{eqnarray}
E_{bw} &=& \int_{r_r}^{r_f} (\Gamma^2 h - p + {\Gamma^2 B^2 \over 4\pi}) 4\pi r^2 dr \nonumber \\
&\approx& 4\pi r_d^2 (\Gamma^2 H - P + {\Gamma^2 {\cal B} \over 4\pi}).
\label{eq:Ebw}
\end{eqnarray}
With the pressure balance assumption, the profile of all the quantities should be uniform in region 2 and region 3, respectively. Therefore, the expression of total energy of the blastwave can be written as\footnote{ Equation \ref{eq:Ebwp} is equivalent to $E_{\rm bw} = {4\pi \over 3}(\Gamma^2 h_r - p_r + {\Gamma^2 B_r^2 \over 4\pi})(r_d^3 - r_r^3) +{4\pi \over 3}(\Gamma^2 h_f - p_f) (r_f^3 - r_d^3)$, when $r_f \sim r_r \sim r_d$. However, with the expansion of the blastwave, the latter equation would introduce an even larger error.}
\begin{eqnarray}
E_{\rm bw} &\approx& 4\pi r_d^2(\Gamma^2 h_r - p_r + {\Gamma^2 B_r^2 \over 4\pi})(r_d - r_r) \nonumber \\
&+& 4\pi r_d^2(\Gamma^2 h_f - p_f) (r_f - r_d).
\label{eq:Ebwp}
\end{eqnarray}
However, both Equation \ref{eq:Ebwp} and the second line of Equation (\ref{eq:Ebw}) are valid only when $r_r \sim r_d \sim r_f$ is satisfied. For the mechanical model, it is convenient to calculate the energy of the blastwave directly through the volume integrals of the quantities, which can reduce the error introduced by spherical expansion. The blastwave energy in the mechanical model reads
\begin{eqnarray}
E_{\rm bw,mech} = \Gamma^2 H_{\rm sph} + P_{\rm sph} + {\Gamma^2 {\cal B}_{\rm sph} \over 4\pi}, 
\label{eq:Ebwm}
\end{eqnarray}
where the volume integrated quantities $H_{\rm sph} = \int_{r_r}^{r_f} 4\pi r^2 h dr$ and $P_{\rm sph} =\int_{r_r}^{r_f} 4\pi r^2 p dr$ can be derived from $H$ and $P$ with the similar relationship shown in Equation \ref{eq:dB}. 

In principle, the total energy of the blastwave should be equal to the total energy injected to the blastwave plus the rest mass energy of the ambient medium being swept ($E_{\rm 1,sw} = {4\pi \over 3}r_f^3 n_1 m_p$).
Thus, the ratio between the two can be used for the energy conservation test. As we can see from the lower right panel of Figure \ref{fig:evolving}, both models satisfy the energy conservation well in the early stage when $r_r \sim r_f$. However, the error increases quickly as the blastwave expands. If the energy of blastwave is calculated with Equation \ref{eq:Ebw} and \ref{eq:Ebwp}. For the pressure balance model, the deviation exceeds $25\%$ within $r_d = 10^{17}{\rm cm}$ with a large $\sigma_{\rm ej}$ values. For the mechanical model, on the other hand, the deviation could be always smaller than $10 \%$ within the distance $r_d = 10^{18}{\rm cm}$ for $\sigma_{\rm ej} < 10$. If the energy of the blastwave is calculated with Equation \ref{eq:Ebwm}, the deviation is negligible within $r_d = 10^{17}{\rm cm}$ and is smaller than $25\%$ within $r = 10^{19}{\rm cm}$ for $\sigma_{\rm ej} < 10$. All in all, the mechanical model satisfies the energy conservation much better than the pressure balance model.

In reality, the central engine timescale cannot be infinitely long. For example, a newly born neutron star  with an initial spin period $P_0 \sim 1~{\rm ms}$ and a fiducial value of moment of inertia $I = 3 \times 10^{45}{\rm erg~s^{-1}}$ would have a total rotational energy $E_{\rm rot} = (1/2)I\Omega^2 \sim 10^{53}{\rm erg}$. Assuming that the magnetized ejecta is the wind of the NS with a luminosity of $L_{\rm ej} = 10^{47}{\rm erg~s^{-1}}$, one can obtain an upper limit of the central engine timescale as $\tau < 10^{6}s$. Since the strength of the poloidal magnetic field decreases with distance from the central NS as $B_p \sim R^{-2}$ while that of the toroidal magnetic field decreases as $B_d \sim R^{-1}$, the magnetic field beyond the light cylinder would be dominated by the $\phi$ component, which agrees with the geometry we discussed in section \ref{sec:evolution}. 

With a finite central engine timescale, the reverse shock would eventually cross the ejecta at some time. Rather than adopting the upper limit of the central engine timescale, here we choose a more realistic value $\tau = 10^{4} s$ as an example\footnote{ For a rapidly spinning NS, there could be other mechanisms (such as secular gravitational waves \citep{fan13,gao16,aloy21}) to release the rotational energy.}. Other parameters are the same as those adopted in Figure \ref{fig:evolving}, thus the evolution history should also be the same. However, instead of always having a stable FS - RS system, there will be a RS crossing time, after which the blastwave would experience a relaxation process before entering the Blandford-McKee regime (the self-similar, asymptotic phase) \citep{blandford76,mimica09} . We stop our calculation at the RS crossing time, when essentially all the energy from the ejecta is injected into the blastwave. Since the rest mass energy of the ambient medium is negligible, the energy of the blastwave should always be the same at this time, regardless of the value of the magnetization parameter $\sigma_{\rm ej}$. 

\begin{figure}
\resizebox{80mm}{!}{\includegraphics[]{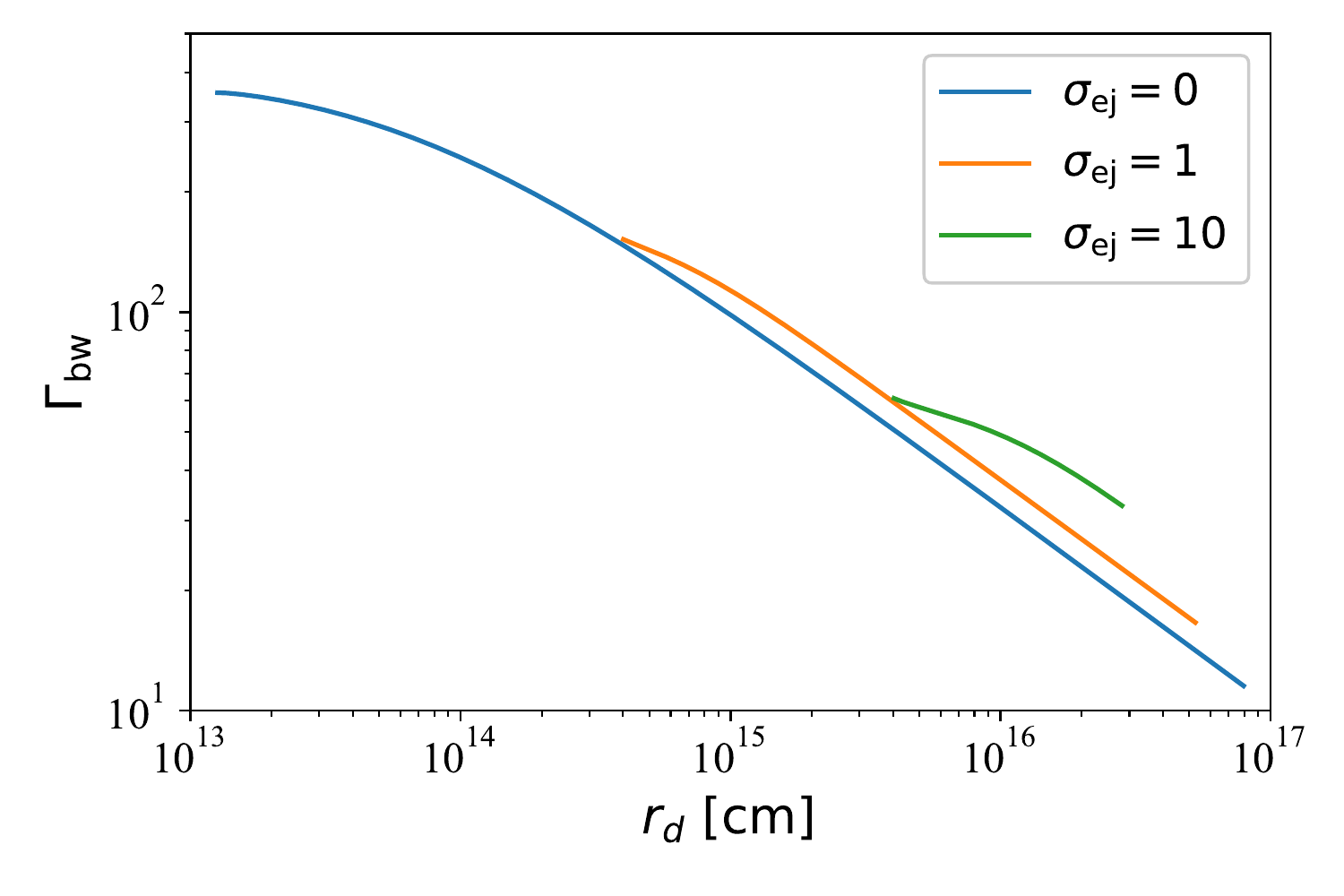}}\\
\resizebox{80mm}{!}{\includegraphics[]{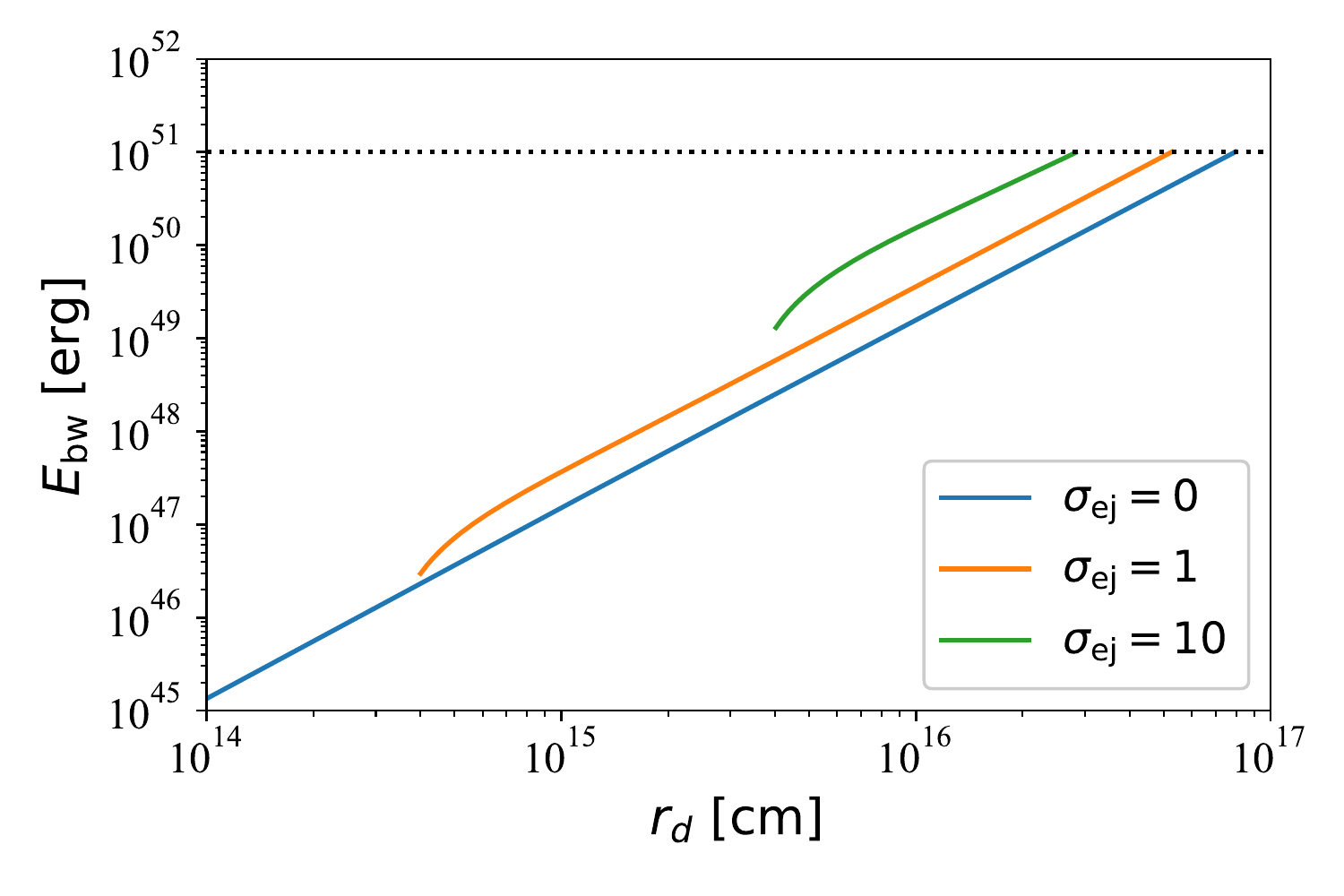}} \\
\resizebox{80mm}{!}{\includegraphics[]{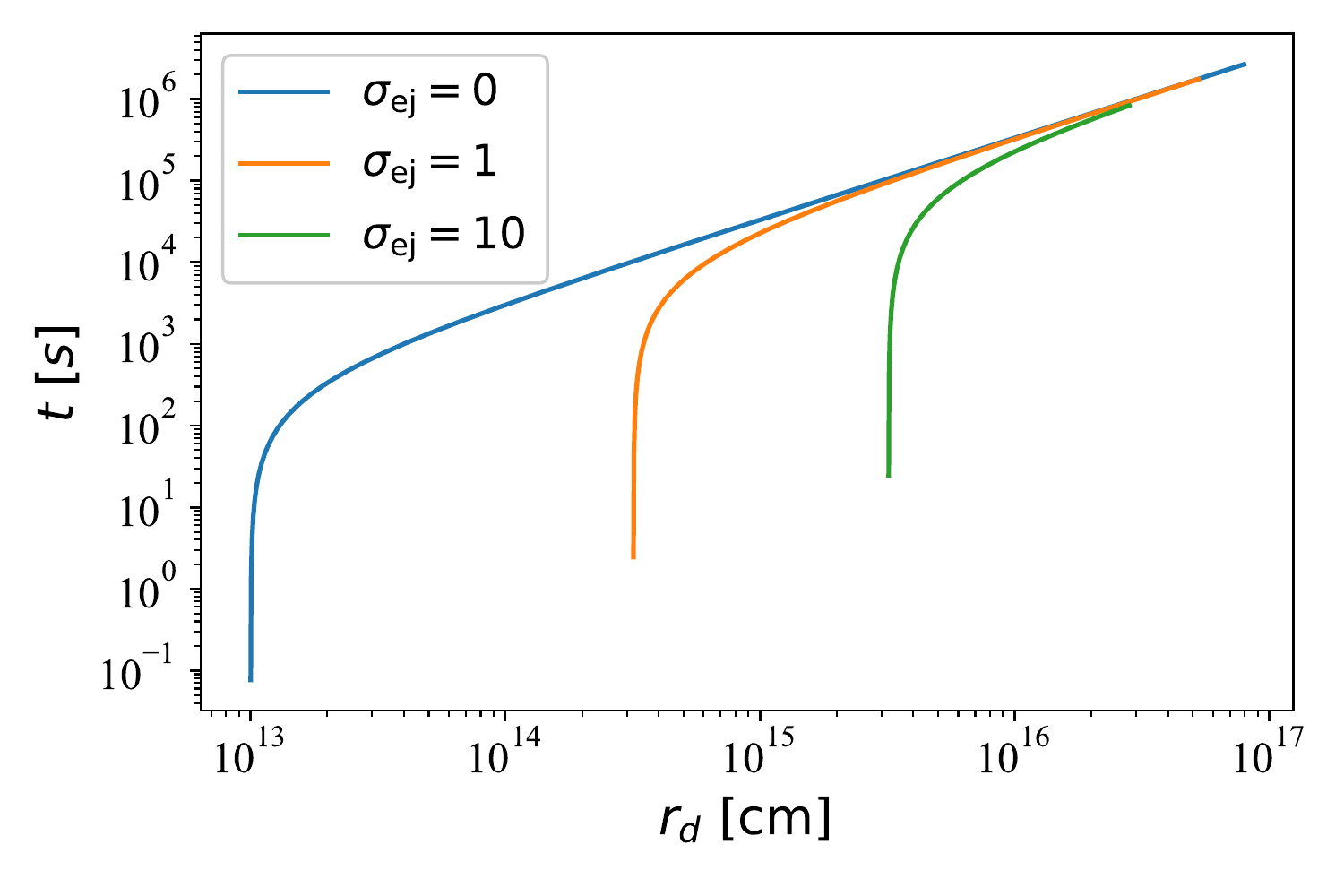}} 
\caption{Blastwave evolution for a central engine with a finite timescale. Upper panel: the evolution of Lorentz factor. Middle panel: the evolution of blastwave energy. Lower panel: the timescale of the evolution of blastwave in the lab frame since the moment when a stable FS - RS system forms. the central engine duration $\tau = 10^4 s$ is adopted for all the panels. Different colors represent different $\sigma_{\rm ej}$ values.}
\label{fig:evolving_tau}
\end{figure}

The results are shown in Figure \ref{fig:evolving_tau}. As we can see, the energy of the blastwave at the RS crossing time is roughly $E_{\rm bw} \sim 10^{51}{\rm erg}$, which is consistent with the value estimated from $E_{\rm bw} \sim L_{\rm ej} \tau$. We also calculate the timescale of the blastwave evolution in the lab frame, which shows that the RS crosses the ejecta earlier for an ejecta with a higher $\sigma_{\rm ej}$. This is understandable since shock propagates faster in the stronger magnetized upstream \citep{zhang05}.

\section{Conclusion and Discussion}\label{sec:con}
In this work, we extended the mechanical model for hydrodynamical blastwaves \citep{beloborodov06,uhm11} to the magnetically dominated regime and calculate the evolution of a blastwave driven by a magnetized ejecta. We break the pressure balance assumption ($p_r = p_f$) and derive the governing equations of the evolution from the basic ideal MHD equations. The blastwave is treated as a whole, i.e. we consider only the integrated quantities of the blastwave rather than the fluid elements and their profiles within the blastwave. By defining four integrated quantities (Equations \ref{eq:Sigma}-\ref{eq:calB}), we derive four govening equations (Equations \ref{eq:mass3} - \ref{eq:energy3}, \ref{eq:eos} and \ref{eq:dB}) to solve the blastwave problem. Through various tests, we find that the mechanical model is in general much better than the pressure balance model in terms of energy conservation, especially in the high $\sigma_{\rm ej}$ regime. The results are also much closer to the numerical simulations results of \cite{mimica09}. For a central engine with an infinitely long central engine time, our mechanical model works precisely at small radii, and only deviates from energy conservation within $25 \%$ for $\sigma_{\rm ej} < 10$ at a distance $r_d < 10^{19}{\rm cm}$ from the central engine. For more realistic cases with limited engine timescale $\tau = 10^4 s$, we checked RS crossing timescales for different $\sigma_{\rm ej}$ values and reached expected results. 

It is worth noticing that the pressure balance treatment is a nice approximation when calculating the evolution of a blastwave with a short engine time or a low $\sigma_{\rm ej}$, so those treatments can give reasonable approximations for GRB FS-RS problems \citep{sari95,zhang05}. However, a mechanical model is needed when dealing with blastwave problems with a long lasting central engine \citep[e.g.][]{uhm12}, especially when the engine is highly magnetized. Our developed theory would be useful to treat problems invoking energy injection of a long-lived engine in various explosive events, including possible pulsar-powered kilonova following neutron star mergers \citep[e.g.][]{yu13,metzger14}.

Another approach of improving the pressure balance model to preserve energy conservation has been discussed in the literature \citep{yan07,chen21}. The energy conservation requirement is imposed by hand, and the pressures are assumed to be uniform in regions 2 and 3 but discontinuous at the contact discontinuity. Such an ad hoc treatment can reach similar conclusion as ours \citep{chen21}, but has larger deviations in the high $\sigma$ regime.

\section*{Acknowledgements}
We thank Miguel-Angel Aloy and Peter Mimica for useful communications, Shao-Ze Li, Liang-Duan Liu, Jing-Ze Ma and Jared Rice for helpful discussion, and an anonymous referee for many helpful comments. This work is supported by the Top Tier Doctoral Graduate Research Assistantship (TTDGRA) at University of Nevada, Las Vegas. 

\section*{Data Availability}
No new data were generated or analysed in support of this research.



%\bibliographystyle{mnras}
%\bibliography{mybib} 

%%%%%%%%%%%%%%%%%%%%%%%%%%%%%%%%%%%%%%%%%%%%%%%%%%

%%%%%%%%%%%%%%%%% APPENDICES %%%%%%%%%%%%%%%%%%%%%

%%%%%%%%%%%%%%%%%%%%%%%%%%%%%%%%%%%%%%%%%%%%%%%%%%

% Don't change these lines
%\bsp	% typesetting comment
\label{lastpage}
\end{document}